\documentclass{ws-ijgmmp}

\begin{document}

\markboth{Hans-Thomas Elze}{A Baker-Campbell-Hausdorff formula for the logarithm of permutations}

%
\catchline{}{}{}{}{}
%

\title{A BAKER-CAMPBELL-HAUSDORFF FORMULA FOR THE LOGARITHM OF PERMUTATIONS}

\author{HANS-THOMAS ELZE}

\address{Dipartimento di Fisica "Enrico Fermi", Universit\`a di Pisa, \\
Largo Pontecorvo 3, I-56127 Pisa, Italia \\ 
elze@df.unipi.it}

\maketitle

\begin{history}
\received{19 December 2019} 
\accepted{28 January 2020}
\end{history}

\begin{abstract}
The dynamics-from-permutations of classical Ising spins is studied for a chain of four spins. We obtain the Hamiltonian operator which is equivalent to the unitary permutation matrix that encodes assumed pairwise exchange interactions. It is shown how this can be summarized by an exact terminating Baker-Campbell-Hausdorff formula, which relates the Hamiltonian to a product of exponentiated two-spin exchange permutations. We briefly comment upon physical motivation and 
implications of this study. 
\end{abstract}


\section{Introduction}	

Our aim in this letter is to extend to the four-spin case the previous study of a chain of three classical Ising spins \cite{ElzeQu19}. 

The analysis of such deterministic systems is motivated by 't\,Hooft's {\it Cellular Automaton Interpretation of Quantum Mechanics} \cite{tHooft2014}. A crucial feature of this is evolution of the underlying {\it Ontological States} ($\cal OS$) by permutations among themselves. 

According to this interpretation of quantum mechanics, the formation of superposition states, which are an unmistakable sign of the quantum theoretical description of what is physically ``out there'', is to be avoided. It only appears in the mathematical description of $\cal OS$ and of their behaviour, {\it cf.} also Ref.\,\cite{Rovelli2015}. 

We do not enter the discussion of important implications for the foundations of quantum theory here -- see Ref.\,\cite{ElzeQu19} and, foremost, the book of Ref.\,\cite{tHooft2014} for this, including numerous references to explorations of various other ``pre-quantum''  models. Instead we presently describe further, especially formal aspects of {\it dynamics-from-permutations}. This will help to prepare the ground for studying most interesting large-$N$ and higher-dimensional systems of this kind.  
  
We describe the proposed dynamics of a composite system of two-state Ising spins by a unitary permutation matrix, which represents the chosen pairwise exchange interactions among $N$ spins. In Section\,2., this is related to so-called 
models of {\it cogwheels with $N$ teeth} \cite{tHooft2014} and we extract the Hamiltonian operator corresponding to their unitary evolution operator. 

These results are employed to obtain the Hamiltonian for the four-spin 
chain, a classical system now described in a manner which closely resembles quantum theory. -- 
It must be emphasized that all models with a finite number of states evolving by permutations can be mapped on suitable combinations of cogwheel models. Yet their physical interpretation rests on further specifying the states and the interactions, eventually including conservation laws.  

Indeed, the extracted Hamiltonian here can be written in a way that looks like a polynomial of spin-exchange operators acting on four-qubit states. In Section\,3., we comment on these  aspects and, furthermore, show that our derivations provide a new {\it Baker-Campbell-Hausdorff formula} incorporating such exchange operators. 

We point out that even small perturbations of the Hamiltonian, generally, will move the dynamical description from being classical, despite its appearance, to becoming the one of a quantum system \cite{ElzeQu19}. Which poses interesting questions left for future work.  

The possible relevance of Ising models for the understanding of Bell's theorem and statistical issues in quantum theory, other than serving as our model for the evolution of $\cal OS$, has recently also been discussed \cite{Vervoort1,Vervoort2,Wetterich16a,Wetterich16b}. 

In Section\,4., we provide a brief summary with our conclusions.  

\section{Simple Facts about Permutations}   

Let $N$ objects, $A_1,A_2,\;\dots \;,A_N$ (``states''), be mapped in $N$ steps onto one another, involving {\it all} states exactly once. 
Such permutations can be represented by {\it unitary} $N\times N$ matrices,  
$\hat U_N$, with exactly {\it one} off-diagonal arbitrary phase 
per column and row, $\exp (i\phi_k), k=1,\;\dots \;,N$, and zero elsewhere. 

Furthermore, by appropriately ordering the states, a given permutation matrix can be put into a 
{\it standard form}, which we choose as follows: 
\begin{equation} \label{UN} 
\hat U_N:= 
\left (
\begin{array}{c c c c c c} 
0           & .           & .           & .     & 0       & e^{i\phi_N} \\ e^{i\phi_1} &0            & .           &       & .       & 0  \\ 
0           & e^{i\phi_2} & 0           & .     &         & .  \\
.           & 0           & e^{i\phi_3} & 0     &         & .  \\ 
.           &             & .           & .     & .       & .  \\
0           & .           & .           & 0     & \;e^{i\phi_{N-1}} & 0  \\
\end{array}\right )
\;\;,\;\;\; 
\hat U_N\hat U_N^\dagger =\mathbf{1} 
\;\;.
\end{equation}

By considering an explicit example, one easily finds that all such  permutation matrices have the property: 
\begin{equation} \label{UNN} 
(\hat U_N)^N=e^{i\sum_{k=1}^N \phi_k}\;\mathbf{1}
\;\;, \end{equation} 
which also follows from the abstract description given above.  
This immediately yields their eigenvalues as the $N$th roots of 1, which lie on the unit circle in the complex plane, multiplied by an overall phase. 	

It is convenient and motivated by the applications we have in mind to define a related {\it Hamiltonian} by: 
\begin{equation} \label{Hop} 
\hat U_N=:e^{-i\hat H_NT} 
\;\;, \end{equation} 
with $T$ introducing a time scale. 
Its eigenvalues, then,  
follow from Eq.\,(\ref{UNN}). This  
results in the diagonal form 
of the Hamiltonian:   
\begin{equation} \label{Hdiag} 
\hat H_{N,\;diag}=\mbox{diag}\Big (\frac{1}{NT}\big (2\pi (n-1)
-\sum_{k=1}^N\phi_k\big )\;|\;n=1,\;\dots\; ,N
\Big ) 
\;\;. \end{equation} 
Different forms of this operator in a variety of contexts will be of central interest in the following. -- Henceforth, we will set the arbitrary phases to zero, $\phi_k=0$, since they will play no role here. 

We point out that these results underlie so-called 
{\it cogwheel models}. They are  
deterministic time reversible models which show quantum mechanical features in suitable limits \cite{tHooft2014,Elze,ElzeRelativPart}.  
For $N\rightarrow\infty$ and $T\rightarrow 0$, 
with $NT\equiv\omega^{-1}$ fixed, the  
{\it quantum harmonic oscillator} can be represented by such a model and for $\omega\rightarrow 0$ a  
{\it free quantum particle}. 

However, difficulties are encountered, when 
one tries to introduce interactions among such 
simplest building blocks, which by themselves represent quantum mechanical one-body systems. Therefore, we initiated to study more complex situations in Ref.\,\cite{ElzeQu19}, namely suitable composites of Ising spins, preparing for future generalizations at present. 

It has been straightforward to find the eigenvalues of permutation matrices, which are incorporated in the diagonal form of the Hamiltonian of Eq.\,(\ref{Hdiag}). 
More generally, however, we also would like to know 
the representation of the Hamiltonian corresponding to the standard form of $\hat U_N$, given in Eq.\,(\ref{UN}).   

We proceed in several steps, as follows. 

\subsection{Eigenvectors, diagonalizing matrix, and equivalent Hamiltonian for a permutation matrix}  

In order to construct the eigenvectors of a permutation matrix, we introduce the {\it auxiliary basis vectors}: 
\begin{equation} \label{AuxVec} 
|\alpha_m\rangle :=(0,\dots ,0,1,0,\dots ,0)^t\;\;,\;
m=1,\dots ,N
\;\;, \end{equation} 
with the entry 1 at the $m$-th position, respectively. The standard form of a permutation matrix $\hat U_N$, {\it cf.} Eq.\,(\ref{UN}), refers to this basis. 
  
Then, we consider an {\it Ansatz} for the normalized eigenvectors, 
$|A_n\rangle ,\; n=1,\;\dots \;,N$, in terms of the auxiliary basis vectors: 
\begin{equation} \label{eigenvec} 
|A_n\rangle =\frac{1}{\sqrt N}\sum_{m=1}^N
e^{ia_{nm}}|\alpha_m\rangle 
\;\;, \end{equation} 
which is suggested by working out a simple example for small $N$ first, such as in Refs.\,\cite{ElzeQu19,tHooft2014}. Momentarily choosing the timescale $T=1$ and inserting the 
Ansatz appropriately into $\hat U_N|A_n\rangle =\exp (-iE_n) |A_n\rangle$, with $E_n=2\pi (n-1)/N$, gives a recursion relation for the phases $a_{nm}$: 
\begin{equation} \label{recursion} 
a_{nm+1}=a_{nm}+E_n\;\;(\mbox{mod}\;2\pi)\;\;,\;\; 
a_{n1}=a_{nN}+E_n\;\;,\;\;a_{n1}:=0 
\;\;. \end{equation}  
This yields: 
\begin{eqnarray} \label{phases} 
a_{nm}&=&\frac{2\pi}{N}(nm-n-m+1)\;\;(\mbox{mod}\;2\pi) 
\\ [1ex] \label{phasessymm}
&=&a_{mn}
\;\;, \end{eqnarray} 
which fully specifies the eigenvectors. 

Furthermore, the phases we obtained also determine the unitary 
{\it diagonalizing matrix} $\hat D$ which maps the auxiliary basis vectors 
to the eigenvectors. This is obvious from the Ansatz (\ref{eigenvec}): 
\begin{equation} \label{diagonalM} 
|A_n\rangle =\sum_{m=1}^N(\hat D)_{nm}|\alpha_m\rangle\;\;,\;\;
(\hat D)_{nm}:=\frac{1}{\sqrt N}e^{ia_{nm}}
\;\;. \end{equation} 
  
With the help of the diagonalizing matrix $\hat D$ we can relate the diagonal form of the Hamiltonian, $\hat H_{N,\;diag}$ given in Eq.\,(\ref{Hdiag}), to the generic form of $\hat H$ defined in Eq.\,(\ref{Hop}): 
\begin{equation} \label{Hgeneric} 
\hat H_N=\hat D^\dagger \hat H_{N,\; diag}\hat D 
\;\;. \end{equation} 
Direct calculation of the matrix elements of $\hat H_N$ then 
gives the result: 
\begin{eqnarray} \label{HgenericM1} 
(\hat H_N)_{nn}&=&\frac{\pi}{NT}(N-1)\;\;,\; n=1,\;\dots\; ,N 
\;\;, \\ [1ex] \label{HgenericM2} 
(\hat H_N)_{n\neq m}&=&\frac{\pi}{NT}
\Big (-1+i\cot\big (\frac{\pi}{N}(n-m)\big ) \Big ) 
\;\;,\; n,m=1,\;\dots\; ,N  
\;\;. \end{eqnarray} 
The Hamiltonian is self-adjoint, as expected. Furthermore, note that the matrix elements are {\it constant} along lines parallel to the diagonal, as well as on the diagonal itself. This implies that the matrix 
elements of adjacent rows or columns differ only by a {\it cyclic permutation}. This observation will turn out to be quite useful.  

At this point, however, the analysis of 
permutations of the standard form given in Eq.\,(\ref{UN}), which represent the {\it cogwheel models} in particular, is completed. At first sight, it appears to leave no room for more interesting physical models, other than  harmonic oscillators, {\it i.e.} free field modes, or free particles. --  
That this is not the case will be shown next.  

\section{Permutations from Bit or Ising Spin Exchanges} 

In Ref.\,\cite{ElzeQu19} we discussed the motivation for and details of the picture of {\it Ontological States} ($\cal OS$) evolving deterministically, which is meant to explain why and how quantum theory successfully represents the reality ``out there'', according to the {\it Cellular Automaton Interpretation of Quantum Mechanics} \cite{tHooft2014}. 

Presently, we do not pursue this discussion of the foundations of quantum theory but wish to 
further explore the potential of {\it dynamics-from-permutations} to give rise to more complex phenomena and some of its interesting formal aspects. 

We have to give a closer look at how permutations are  introduced and realized with respect to auxiliary basis vectors, {\it cf.} Eqs.\,(\ref{UN}) and (\ref{AuxVec}). 
Of course, the dimension of the $N\times N$-matrix $\hat U_N$ can be varied, but the formulation is general and applies equally well for small systems as for systems with $N>>1$. 

Yet, so far, we have said nothing in more physical terms about {\it how $\cal OS$ are built and how come they evolve by permutations} among themselves. In the following, we will do this by introducing a suitable classical spin model.   

\subsection{Exchange operations and permutations} 
 
Let us consider a chain which consists of four classical {\it two-state Ising spins}, labeled ``1,2,3,4'', which are equivalent to four Boolean variables or four bits.\footnote{We considered the 
three-spin case in Ref.\,\cite{ElzeQu19} and intend to report on the $N$-spin case elsewhere.} They can be in one of $2^4=16$ states.\footnote{Assuming such multi-spin states as $\cal OS$, there are {\it no superposition states}, by definition \cite{tHooft2014}.}  

A coupling among the four spins, leading to permutations among the 16 
states, can be generated 
by {\it spin exchange}, a permutation 
$\hat P_{ij}\;(\equiv\hat P_{ji})$ involving two spins, labeled $i,j=1,\;\dots\;,4$, with the following properties: 
\begin{equation} \label{Pijdef} 
\hat P_{ij}|s_i,s_j\rangle :=
|s_j,s_i\rangle 
\;\;,\;\;\; \hat P_{ji}\hat P_{ij}
=(\hat P_{ij})^2=\mathbf{1}
\;\;, \end{equation} 
with the states of a single spin denoted by   
$s_k=\pm 1$ or, equivalently, by $s_k=\uparrow ,\downarrow$, for ``spin up, spin down'', respectively; we use the ket notation $|s_i,s_j\rangle$ 
to indicate that the first spin has value $s_i$, the second value $s_j$, and analogously for a state of four spins. 

An obvious property of all permutations is that the {\it numbers of up and down spins are separately conserved}:  
\begin{equation} \label{Nupdown} 
[\hat N_u,\hat P_{ij}]=[\hat N_d,\hat P_{ij}]=0\;\;,\;\mbox{for all}\; i,j 
\;\;, \end{equation} 
with the number operators defined explicitly below, making use of the Pauli matrix $\hat\sigma^z$. -- 
This implies that not all of the $2^4$ states can be reached from a given initial state, no matter how we define the dynamics in terms of exchange operators. 

Furthermore, for any state with $N_u$ up- and $N_d$ down-spins and for all sequences of permutations acting on it, there is a state with $N_d$ up- and $N_u$ down-spins which evolves under these permutations in one-to-one correspondence with the former one -- {\it i.e.}, the {\it spinflip  operator}, symbolically $\hat C:\;\uparrow\leftrightarrow\downarrow$,  commutes especially with the exchange operations:
\begin{equation} \label{C} 
[\hat C,\hat P_{ij}]=0\;\;,\;\mbox{for all}\; i,j 
\;\;. \end{equation} 
Also $\hat C$ can be defined in terms of Pauli matrices. 

In fact, we may identify the two states $s_k=\pm 1$ of an Ising spin with the eigenstates of the 
Pauli matrix $\hat\sigma^z$, 
$\psi_+=(1,0)^t$ and $\psi_-=(0,1)^t$, respectively. Then, the unitary operator 
$\hat P_{ij}$ can be expressed in terms 
of the Pauli spin-1/2 matrices: 
\begin{equation} \label{PijPauli}
\hat P_{ij}=\frac{1}{2}(\underline{\hat\sigma}_i
\cdot\underline{\hat\sigma}_j+\mathbf{1})
\;\;, \end{equation} 
where $\underline{\hat\sigma}$ is   
the vector with components  $\hat\sigma^x,\hat\sigma^y,\hat\sigma^z$. 
The above mentioned number operators are $\hat N_u:=2+\sum_{i=1}^4\hat\sigma_i^z/2$ and $\hat N_d:=4-\hat N_u$. Furthermore, a total spinflip is effected by 
$\hat C:=\prod_{i=1}^4\hat\sigma_i^x$. 
These expressions hint towards a relation with QM of qubits, which we discussed before \cite{ElzeQu19} and shall briefly recall in the following.

Finally, two permutations involving three different spins do not commute: 
\begin{equation} \label{comm} 
[\hat P_{ij},\hat P_{jk}]\neq 0\;\;, \;\;
\mbox{for}\;\; i\neq k
\;\;, \end{equation} 
with no summation over $j$ implied.\footnote{With $a,b,c=\pm 1$, arbitrary but fixed,  
we have, {\it e.g.}, $\hat P_{12}\hat P_{23}|abc\rangle =|cab\rangle\neq |bca\rangle =
\hat P_{23}\hat P_{12}|abc\rangle$.}  
This {\it noncommutativity} is the reason that extracting a Hamiltonian from a unitary update operator, now acting on a collection of classical spin variables by permutations, needs some additional steps as compared to what has been done in Section\,1.   

\subsection{Unitary evolution} 

We have to choose a unitary operator $\hat U$ that evolves states of four classical spins in a discrete time step $T$ 
and our aim is to extract the corresponding Hamiltonian $\hat H$, 
{\it cf.} Eqs.\,(\ref{Hop})-(\ref{Hdiag}).  
Let us consider: 
\begin{equation} \label{U} 
\hat U:=\hat P_{23}\hat P_{12}\hat P_{34}=:\exp (-i\hat HT) 
\;\;, \end{equation} 
acting sequentially on the indicated pairs of spins. The analysis for any other choice of the sequence of three exchange operators would proceed analogously to the following, provided all spins are addressed at least once. 

Because of the {\it conservation laws}, Eqs.\,(\ref{Nupdown}) and (\ref{C}), 
we expect $\hat U$ to have a block diagonal structure. By inspection we find that the 16 states of the four spins can be meaningfully grouped:  
\begin{equation} \label{states116} 
|1\rangle :=|\uparrow ,\uparrow ,\uparrow,\uparrow\rangle 
 =\hat U|1\rangle \;\;,\;\; 
|16\rangle :=\hat C|1\rangle =|\downarrow ,\downarrow ,\downarrow,\downarrow\rangle =\hat U|16\rangle 
\;\;, \end{equation} 
\begin{eqnarray}  
|2\rangle :=|\uparrow ,\uparrow ,\uparrow ,\downarrow\rangle 
 =\hat U|5\rangle \;&,&\;\; 
|12\rangle :=\hat C|2\rangle
\;\;, \nonumber \\ 
|3\rangle :=|\uparrow ,\downarrow ,\uparrow ,\uparrow\rangle 
 =\hat U|2\rangle \;&,&\;\; 
|13\rangle :=\hat C|3\rangle
\;\;, \nonumber \\
|4\rangle :=|\downarrow ,\uparrow ,\uparrow ,\uparrow\rangle 
 =\hat U|3\rangle \;&,&\;\; 
|14\rangle :=\hat C|4\rangle
\;\;, \nonumber \\ \label{states25}
|5\rangle :=|\uparrow ,\uparrow ,\downarrow ,\uparrow\rangle 
 =\hat U|4\rangle \;&,&\;\;
|15\rangle :=\hat C|5\rangle 
\;\;, \end{eqnarray} 
\begin{eqnarray}   
|6\rangle :=|\uparrow ,\uparrow ,\downarrow ,\downarrow\rangle 
=\hat U|9\rangle \;&,&
\nonumber \\ 
|7\rangle :=|\uparrow ,\downarrow ,\uparrow ,\downarrow\rangle 
=\hat U|6\rangle \;&,&
\nonumber \\
|8\rangle :=|\downarrow ,\downarrow ,\uparrow ,\uparrow\rangle 
=\hat U|7\rangle \;&,&
\nonumber \\ \label{states69}
|9\rangle :=|\downarrow ,\uparrow ,\downarrow ,\uparrow\rangle 
=\hat U|8\rangle \;&,& 
\end{eqnarray} 
\begin{equation} \label{states1011}
|10\rangle :=|\downarrow ,\uparrow ,\uparrow ,\downarrow\rangle 
=\hat U|11\rangle \;\;,\;\; 
|11\rangle :=\hat C|10\rangle 
=|\uparrow ,\downarrow ,\downarrow ,\uparrow\rangle 
=\hat U|10\rangle 
\;\;. \end{equation} 
Thus, we observe that there are two eigenstates, 
Eqs.\,(\ref{states116}), three groups of four states each that evolve periodically among each other, Eqs.\,(\ref{states25})-(\ref{states69}), and two states that flip-flop between each other, Eqs.\,(\ref{states1011}). 

Correspondingly, the evolution operator $\hat U$ as well as the Hamiltonian $\hat H$ of Eq.\,(\ref{U}) indeed have block diagonal structure. In order to determine the Hamiltonian, we proceed as follows. 

We consider one of the $4\times 4$ blocks, for example, concerning the states $|2\rangle ,\;\dots\; ,|5\rangle$ of Eqs.\,(\ref{states25}). They evolve into each other consecutively, like four consecutive members of the auxiliary basis defined in 
Eq.\,(\ref{AuxVec}) evolve into each other under $\hat U_4$, {\it cf.} Eq.\,(\ref{UN}) (with zero phases). Hence, we know the relevant Hamiltonian with respect to the {\it auxiliary basis} from the results of Section\,2.1., Eqs.\,(\ref{Hgeneric})-(\ref{HgenericM1}): 
\begin{equation} \label{H4} 
\hat H_4=\frac{3\pi}{4T}
\left (\begin{array}{c c c c} 
1   & c^* & d   & c   \\ 
c   & 1   & c^* & d   \\ 
d   & c   & 1   & c^* \\ 
c^* & d   & c   & 1   \\ 
\end{array}\right ) 
\;\;,\;\;\;\mbox{with}\;\; c:=\frac{1}{3}(-1+i)\;\;,\;\; d:=-\frac{1}{3}
\;\;. \end{equation} 

By identifying the four auxiliary basis vectors, to which 
$\hat H_4$ refers, with the four Ising spin states of Eqs.\,(\ref{states25}), $|2\rangle ,\;\dots\; ,|5\rangle$, we can rewrite this Hamiltonian in terms of operators acting on these equivalently evolving {\it four-spin states}. We obtain: 
\begin{eqnarray} \label{H4final} 
\hat H_4&=&\frac{3\pi}{4T} 
\left (\mathbf{1}+c\hat U+d\hat U^2+c^*\hat U^3
\right) 
\nonumber \\ [1ex] 
&=&\frac{3\pi}{4T} 
\left (\mathbf{1}+c\hat U+c^*\hat U^\dagger +d\hat U^2
\right) 
\nonumber \\ [1ex] 
&=&\frac{3\pi}{4T} 
\left (\mathbf{1}+c\hat P_{23}\hat P_{12}\hat P_{34}
+c^*\hat P_{34}\hat P_{12}\hat P_{23} 
+d\hat P_{23}\hat P_{14}
\right) 
\;\;, \end{eqnarray} 
with $\hat U$ of Eq.\,(\ref{U}), using $\hat U^3=\hat U^\dagger$, which follows from unitarity and $\hat U^4=\mathbf{1}$, and calculating explicitly 
$\hat U^2=\hat P_{23}\hat P_{14}$.   

This result for $\hat H_4$ is obviously also correct for the other two $4\times 4$ blocks to be considered. It furthermore gives the correct 
zero energy eigenvalues for the two constant states of Eqs.\,(\ref{states116}), since $1+c+c^*+d=0$. 

However, somewhat more surprisingly, it also describes correctly the behaviour of the two flip-flop states $|10\rangle ,|11\rangle$ of Eqs.\,(\ref{states1011}). On these states {\it only}, we have that $\hat U\equiv\hat U^\dagger$ and $\hat U^2\equiv\mathbf{1}$. Thus, we find here from the second line in Eqs.\,(\ref{H4final}): 
\begin{equation} \label{Hflipflop} 
\hat H_4
=\frac{\pi}{2T}  
\left (\mathbf{1}-\hat U
\right) \;\;\;\longleftrightarrow\;\;\; 
\hat H_2=\frac{\pi}{2T}\left (
\begin{array}{c c} 
1   & -1 \\ 
-1  & 1  \\ 
\end{array}
\right )
\;\;, \end{equation} 
where the Hamiltonian on the left refers to the four-spin states $|10\rangle ,|11\rangle$, while the one on the right refers to two equivalently evolving auxiliary basis states, as determined by  Eqs.\,(\ref{HgenericM1})-(\ref{HgenericM2}).   
  
In conclusion, we have obtained the Hamiltonian $\hat H$ defined by the 
unitary permutation matrix $\hat U$, Eq.\,(\ref{U}), 
which acts as discrete update operator on four-spin states pertaining to a classical Ising model. It is uniformly given by: 
\begin{equation} \label{Hfinal} 
\hat H=\frac{3\pi}{4T} 
\left (\mathbf{1}+c\hat P_{23}\hat P_{12}\hat P_{34}
+c^*\hat P_{34}\hat P_{12}\hat P_{23} 
+d\hat P_{23}\hat P_{14}
\right ) 
\;\;, \end{equation} 
a quite simple result indeed, where the constants $c,d$ are as in Eqs.\,(\ref{H4}). Note that the terms 
on the right-hand side {\it all commute} with each other, since they are given by powers of $\hat U$. 

Since the eigenvalues 
of $\hat H$ are essentially given by the eigenvalues of 
$\hat H_4$, appearing three times, and of $\hat H_2$, see Eq.\,(\ref{Hdiag}), plus two additional zero eigenvalues, we observe a highly {\it degenerate spectrum}, caused by the block diagonal structure of the evolution operator. Physical consequences of this shall be discussed together with a generalization for $N$ spins elsewhere.  

\subsection{Another kind of Baker-Campbell-Hausdorff formula} 
The results we have obtained, in particular the derivation of the self-adjoint Hamiltonian that corresponds to the {\it logarithm of 
a permutation} acting on states of multiple classical spins, 
$$-i\hat HT=\log (\hat P_{23}\hat P_{12}\hat P_{34}) 
\;\;, $$ 
can be nicely summarized in the form of a 
new {\it Baker-Campbell-Hausdorff} (BCH) {\it formula}. 

The related algebraic problem concerning noncommuting operators is familiar from QM or Lie group theory. -- Consider 
$\exp (X)\exp (Y)=\exp (Z)$. The formal solution for $Z$ in terms of $X,Y$ is provided by the BCH formula:   
\begin{equation} \label{BCH} 
Z=X+Y+\frac{1}{2}[X,Y]+\frac{1}{12}
([X,[X,Y]]+[Y,[Y,X]])  
-\frac{1}{24}[Y,[X,[X,Y]]]
+\;\dots
\;\;, \end{equation}  
which presents a series of increasingly complicated iterated commutators. Its coefficients are known. Nevertheless, convergence or divergence of the series is difficult to assess, in general. There are exceptional cases, however, when the series terminates. In recent works by Visser {\it et al.} and by Matone, 
with numerous references to earlier work, some new classes of such closed form solutions have been derived \cite{Visser,Matone}.    

Returning to Eq.\,(\ref{U}), we see that the permutations composing the evolution operator $\hat U$ can be exponentiated separately as follows: 
\begin{equation} \label{exp} 
\hat P_{ij}=i\exp (-i\frac{\pi}{2}\hat P_{ij}) 
\;\;, \end{equation} 
using $(\hat P_{ij})^2=\mathbf{1}$, {\it cf.} Eqs.\,(\ref{Pijdef}). However, since 
$[\hat P_{23},\hat P_{12}]\neq 0$ and $[\hat P_{23},\hat P_{34}]\neq 0$, we cannot evaluate 
the Hamiltonian $\hat H$ by simply adding the exponents 
obtained in this way. 

Instead, we may rewrite Eq.\,(\ref{U}) in a variety of different ways that respectively represent one terminating BCH formula: 
\begin{eqnarray} 
&\phantom .&i^3\exp \big (-i\frac{\pi}{2}\hat P_{23}\big ) 
\exp \big (-i\frac{\pi}{2}\hat P_{12}\big )
\exp \big (-i\frac{\pi}{2}\hat P_{34}\big ) 
\nonumber \\ [1ex] \label{BCH1} 
&=&i^3\exp \big (-i\frac{\pi}{2}\hat P_{23}\big ) 
\exp \big (-i\frac{\pi}{2}(\hat P_{12}+\hat P_{34})\big ) 
\\ [1ex] \label{BCH2} 
&=&i^2\exp \big (-i\frac{\pi}{2}\hat P_{23}\big ) 
\exp \big (-i\frac{\pi}{2}\hat P_{12}\hat P_{34}\big ) 
\\ [1ex] \label{BCH3} 
&=&\exp\big ( 
-i\frac{3\pi}{4} 
(\mathbf{1}+c\hat P_{23}\hat P_{12}\hat P_{34}
+c^*\hat P_{34}\hat P_{12}\hat P_{23} 
+d\hat P_{23}\hat P_{14})\big ) 
\;\;, \end{eqnarray}
where we make use of Eqs.\,(\ref{Hfinal}) and (\ref{exp}),  with $c:=\frac{1}{3}(-1+i)$, $d:=-\frac{1}{3}$, as before.  
The equality of the right-hand sides of 
Eqs.\,(\ref{BCH1}) and (\ref{BCH2}) 
can be demonstrated by series expansions of the 
differing exponentials, taking into account that always 
$(\hat P_{ij})^2=\mathbf{1}$ and particularly 
$[\hat P_{12},\hat P_{34}]=0$. 
Furthermore, since all terms which stem from the Hamiltonian commute, the right-hand side of (\ref{BCH3}) can also be factorized into a product of exponentials, when needed. 

Note that the explicit ``coupling'' constant 
$\pi /2$ can be replaced in the above formulae case by case by $(2k+1/2)\pi$, with integer $k$. This does not affect the validity of the respective formula, due to the same property in Eq.\,(\ref{exp}). Similarly, if instead $\pi /2$ is replaced by $(2k+3/2)\pi$, then an overall sign change has to be included in each case. 

In Ref.\,\cite{ElzeQu19}, we discussed further the 
implications of analogous results for a classical three-spin system. -- 
Namely, if we consider the classical Ising spins as embedded into the larger Hilbert space of qubits, then the Hamiltonian $\hat H$ of Eq.\,(\ref{Hfinal}) expressed in terms of appropriate Pauli matrices, 
{\it cf.} Eq.\,(\ref{PijPauli}), appears {\it as if} being of genuinely quantum mechanical kind. 

In this way, there arises a {\it ``quantum instability''} of the classical spin system:  {\it small perturbations} of the fixed ``coupling'' constants turn the Hamiltonian into an operator that produces superpositions of multi-spin states, a hallmark of quantum mechanics. Whereas the exact Hamiltonian, as above, produces permutations among {\it Ontological States}, by construction.  

\section{Conclusions} 

Motivated by recent studies of ontological models illustrating the {\it Cellular Automaton Interpretation of Quantum Mechanics}, 
we analyze a chain of four classical Ising spins which evolve deterministically by specifically chosen permutations among its $2^4$ states. 

We first present the $N$-cogwheel model in Section\,2., which describes a system that periodically evolves through a discrete set of $N$ states  \cite{tHooft2014}. 
This is related to properties of a unitary $N\times N$ {\it permutation matrix in standard form}. Considering this as update operator for the evolution of the system in a finite time step, we  
obtain the corresponding Hamiltonian operator, its eigenvalues, and its eigenstates. 

Since this model does not require any additional information on the nature of the states nor on their dynamics, besides resulting from permutations, the results can find varied applications. 

Thus, in Section\,3., it is shown that the  four-spin chain, with dynamics generated by permutations based on Ising spin (or {\it bit}) exchanges, can be related to 
a composition of three cogwheels with four teeth each, {\it i.e.} four states, plus two flip-flop states and two additional time independent states. Therefore, the results of Section\,2. can be applied to extract its Hamiltonian, resulting in Eq.\,(\ref{Hfinal}). It is characterized by a highly degenerate spectrum, which is related to conservation laws that cause the evolution operator to have a block diagonal structure.   

Next, {\it via} the exponential of the Hamiltonian and its preceding definition through permutations in the form of spin exchanges, our results are concisely rewritten as a new {\it Baker-Campbell-Hausdorff formula} with terminating expansion in terms of permutation operators, {\it cf.} Eqs.\,(\ref{BCH})-(\ref{BCH3}). 

While the short Ising spin chain treated here is an intrinsically classical deterministic system, we have pointed out the possibility 
of a {\it ``quantum instability''}, which is visible in the BCH formula: small perturbations can spoil this relation and will, generally, lead to dynamics that produces superposition states of four {\it qubits}. Thus, it would replace the underlying classical ontological features by what appears ``naturally'' described by quantum theory. 

It will be most interesting to see the generalization of our approach for large systems ($N\gg 1$), which is under study. One may be curious about situations which are not described by one-dimensional chain-like structures. Last not least, new features relating ontological {\it dynamics-from-permutations} to a quantum mechanical description can be expected to arise from coarse-graining for large systems. 

\section*{Acknowledgments} 
It is a pleasure to thank M. Blasone, A. Elitzur, G. 't Hooft, A. Khrennikov,  G. Vitiello, and C. Wetterich for continuing discussions. Related work has been presented in the session on 
{\it Foundations of Quantum Theory} at PAFT2019: {\it Current Problems 
in Theoretical Physics} (Vietri sul Mare, April 13-17, 2019). Special thanks to the organizers for the invitation to convene this session.

\end{document}